\documentclass{main}

\begin{document}
    \title{FGGM: Formal Grey-box Gradient Method for Attacking DRL-based MU-MIMO Scheduler}
\author{
    \IEEEauthorblockN{
        Thanh~Le\textsuperscript{1}\textsuperscript{2},
        Hai~Duong\textsuperscript{3},
        Yusheng~Ji\textsuperscript{2}\textsuperscript{1}, 
        ThanhVu~Nguyen\textsuperscript{3}
        and
        John C.S.~Lui\textsuperscript{4}
    }
    \IEEEauthorblockA{
        \textsuperscript{1}The Graduate University for Advanced Studies (SOKENDAI), Japan
    }
    \IEEEauthorblockA{   
        \textsuperscript{2}National Institute of Informatics (NII), Japan
    }
    \IEEEauthorblockA{
        \textsuperscript{3} George Mason University, USA
    }
    \IEEEauthorblockA{
        \textsuperscript{4}The Chinese University of Hong Kong, China
    }
}

    \maketitle
    \begin{abstract}
    In 5G mobile communication systems, multi-user multiple-input multiple-output (MU-MIMO) has been applied as a core technology to enhance spectral efficiency and support high data rates for accommodating multiple users.
    To fully exploit the spatial diversity and maximize spectral efficiency while providing fairness among users, the base station (BS) needs to selects a subset of users for data transmission.
    Given that this problem is NP-hard, deep reinforcement learning (DRL)-based methods have been proposed to infer the near-optimal solutions in real-time~\cite{an2023deep}.
    However, this approach has an ``\emph{intrinsic security problem}''.
    This paper investigates how a group of adversarial users can exploit unsanitized raw CSIs to launch a throughput degradation attack.
    We assume that the group of adversarial users have obtained the DRL policy.
    Most existing studies only focused on systems in which adversarial users can obtain the exact values of victims' CSIs, but this is impractical in the case of uplink transmission in LTE/5G mobile systems.
    We note that the DRL policy contains an observation normalizer which has the mean and variance of the observation to improve training convergence.
    Adversarial users can then estimate the upper and lower bounds of the local observations including the CSIs of victims based solely on that observation normalizer.
    We develop an attacking scheme \gls{fggm} by leveraging polytope abstract domains, a technique used to bound the outputs of a neural network given the input ranges.
    Our goal is to find one set of intentionally manipulated CSIs which can achieve the attacking goals for the whole range of local observations of victims.
    Experimental results demonstrate that \gls{fggm} can determine a set of adversarial CSI vector controlled by adversarial users, then reuse those CSIs throughout the simulation to reduce the network throughput of a victim up to 70\% without knowing the exact value of victims' local observations.
    This study serves as a case study and can be applied to many other DRL-based problems, such as a knapsack-oriented resource allocation problems.
\end{abstract}

\begin{IEEEkeywords}
    deep reinforcement learning, neural networks verification, adversarial attack, multi-user massive MIMO
\end{IEEEkeywords}

    \glsresetall
    \section{Introduction}
\label{sec:introduction}
    Massive \gls{mumimo} is a core technology that can greatly enhance the efficiency of current access networks by using numerous antennas at the base station \cite{marzetta2016fundamentals}.
    This allows the base station to perform multi-user beamforming and to serve many user in the same time-frequency resource block.
    However, in dense environments where users are closely located, significant inter-user interference can occur.
    This issue can be mitigated by selecting only a group of users that contribute to the highest overall spectral efficiency and fairness \cite{mounika2021downlink, lau2005proportional}.

    Since \gls{csi} between users and base station change rapidly in mobile networks where users move frequently, as a result, base station must resolve the fair resource scheduling problem while maximizing spectral efficiency in real-time \cite{le2022tinyqmix,le2024mftts}.
    However, the fair scheduling problem can be reformulated as an \gls{ilp} problem and is NP-hard \cite{lau2005proportional}.
    The complexity of solving \gls{ilp} makes it challenging to design optimal yet fast schedulers for next-generation access networks, especially with a large number of users and resource blocks.
    Recent studies applied \gls{drl} to find near-optimal solutions for NP-hard problems such as \gls{tsp} or knapsack problems \cite{bello2016neural}.
    In \cite{an2023deep}, \gls{drl}-based \gls{mumimo} schedulers were trained and tested on realistic channel models and real-world channel measurements, and demonstrated a close performance to the \gls{optpf} scheduler in terms of spectral efficiency and fairness.
    Especially, the \gls{drl}-based scheduler has a small computational complexity, so it can solve the problem in network scenarios where \gls{optpf} is computationally infeasible.

    Nevertheless, the \gls{mumimo} scheduler based on \gls{drl} introduces an ``\emph{intrinsic security problem}'' as it utilizes the unsanitized \gls{csi} from users to make scheduling decisions.
    \gls{drl} is known to be susceptible to adversarial observations crafted by attackers to manipulate the decisions of \gls{drl} agents, which can significantly degrade the performance of a well-trained \gls{drl} policy \cite{huang2017adversarial}.
    Given that the reported \gls{csi}s form part of the inputs/observations for the \gls{drl}-based scheduling policy, a few adversarial users can falsify their \gls{csi}s to negatively influence user selection outcomes such as to degrade the throughput of normal users, e.g., victims.

    In this work, we investigate how to craft the most detrimental adversarial \gls{csi}s, which are determined by the trained \gls{drl} policy and the \gls{csi}s of victims.
    We propose a threat model in which adversaries know the trained \gls{drl} policy for white-box testing, as it can generate better adversarial examples compared to black-box testing.
    This represents the worst-case scenario that network operators need to prepare for.
    If the \gls{drl}-based scheduler can survive this scenario, it would have a better chance of being robust against attacks that are agnostic to \gls{drl} policies.
    
    Most prior works on attacking \gls{mumimo} resource allocation algorithms relied on precise knowledge of victims' \gls{csi}, which adversarial users can capture by listening to the plain text \gls{csi} feedback to the base station \cite{hou2022muster,manoj2021adversarial}.
    This approach might be impractical for systems with uplink \gls{csi} reference signal, where the base station estimates \gls{csi} instead of relying on users feedback, or for systems with physical layer encryption of \gls{csi} feedback.
    On the other hand, our proposed threat model assumes that the adversaries work for all potential values within the normal ranges of victim observations including the victim \gls{csi}s. 
    The total domain of the observation space is established by concatenating the adversarial users' concrete observations with the lower and upper bounds of the victims' observations.
    Our conceptual contribution is the grey-box threat model for \gls{drl}-based \gls{mumimo} scheduler, which incorporates both white-box (\gls{drl} policy) and black-box testing elements (victims' observations).

    We then propose the \gls{fggm} to solve for the adversarial \gls{csi} under our grey-box threat model. 
    \gls{fggm} leverages the polytope abstraction domain \cite{singh2019abstract}, which is commonly used in \gls{dnn} verification \cite{liu2021algorithms,duong2023dpll}, to handle the observation intervals. 
    \gls{fggm} subsequently propagates these observation boundaries through the hidden and output layers of the \gls{drl} policy to construct an objective function on the output bounds.
    Note that the upper bounds of the output determine the chance that base station select an user, i.e. the highest possible performance when base station selecting that user.
    We formulate the problem of finding adversarial \gls{csi}s such that the upper bounds for all outputs that allow victims to transmit data are minimized for all possible victim \gls{csi}s.
    Thus, this attack strategy provides \emph{``a single adversarial input''} at the initiation of the attack, without the need of recalculating during the attack.
    Given that the abstract domain is differentiable, we solve this problem using stochastic gradient descent.

    To this end, this paper introduces a new and practical adversarial attack strategy for \gls{drl}-based massive \gls{mumimo} schedulers.
    The key contributions of this paper are:
    \begin{enumerate}
        \item Develop a grey-box threat model for \gls{drl}-based \gls{mumimo} schedulers, which encompasses the trained \gls{drl} model and possible ranges of victims' observations.
        \item Propose \gls{fggm} to solve for adversaries' observations that degrade victim throughput for all victim observations.
        \item Demonstrate that \gls{fggm} outperforms black-box or sampling-based white-box attacks via \gls{pgd} in reducing access probability and degrading throughput across various widths of input bounds and search spaces.
        \item A single attacker can degrade victim throughput by up to 9.5\%. When 50\% of users are attackers, the victim's throughput can be drastically reduced by up to 70\%.
        \item The implementation of \gls{fggm} is publicly available at \texttt{(redacted)}.
    \end{enumerate}

    This paper is organized as follows. 
    Section \autoref{sec:preliminary} presents the target \gls{drl}-based \gls{mumimo} scheduler and gives a primal on \gls{dnn} verification. 
    Section \autoref{sec:threat_model} explains grey-box threat model, which envisions the capabilities and limitations of the attackers. 
    We describe our proposed \gls{fggm} attack in Section \autoref{sec:propose_attack}, followed by the evaluation and discussion section.   
    \section{Preliminaries}
    \label{sec:preliminary}
    In this section, we briefly describe the target \gls{mumimo} network model, explain why \gls{mumimo} scheduling is a difficult problem, and how it was addressed by using \gls{drl}.
    We then describe how \gls{drl}-based schedulers which can infer the near-optimal solution for this problem in real-time.

    \subsection{Network Models}
        \begin{figure}[ht]
            \centering
            \includegraphics[width=\linewidth]{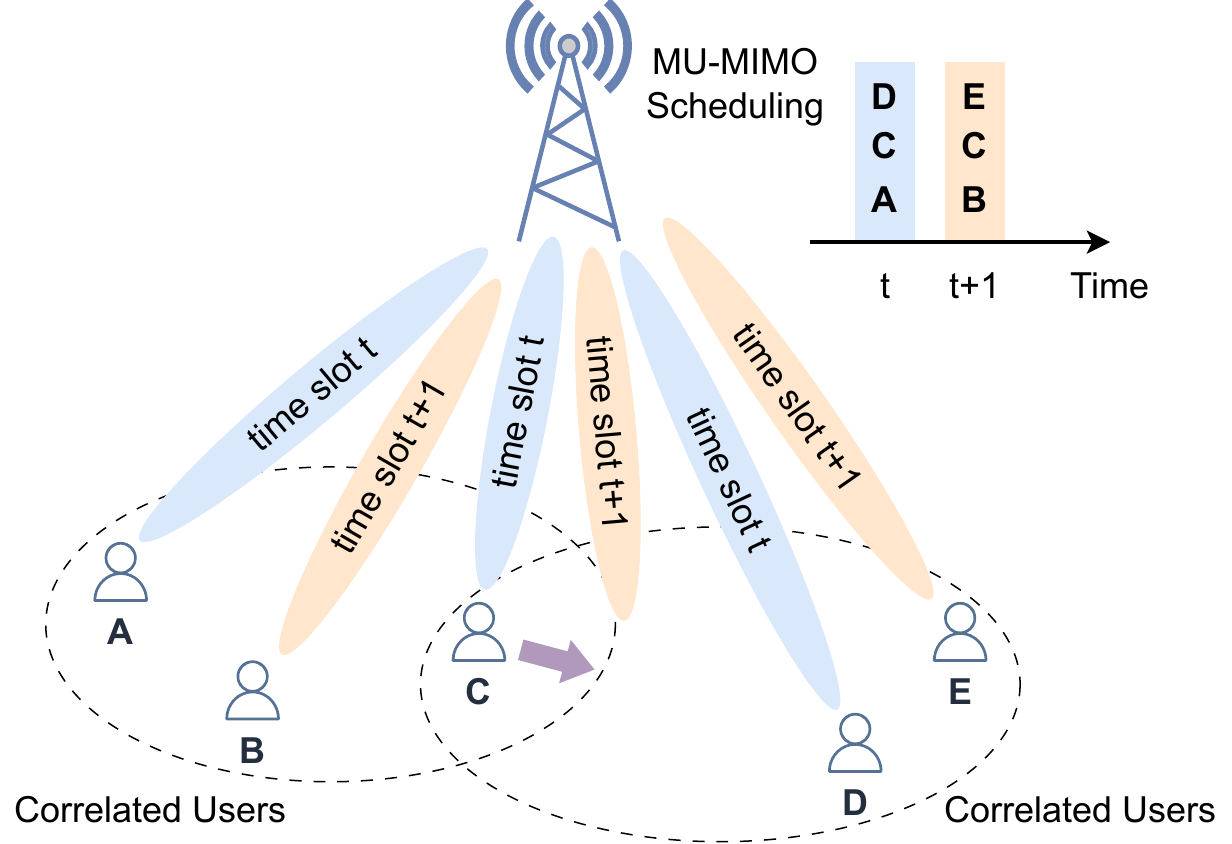}
            \caption{The system model of the \gls{mumimo} scheduling problem.}
            \label{fig:preliminary:system_model}
        \end{figure}

        Consider a single-cell wireless network base station with $M$ \gls{mimo} antennas (\autoref{fig:preliminary:system_model}).
        The base station uses the \gls{ofdm} technique with $B$ orthogonal resource blocks, and performs \gls{mumimo} transmission in a time-slotted manner.
        Each resource block is shared among $L$ single-antenna mobile user.
        Assume that the \gls{tdd} mode is used, where $L$ users periodically send orthogonal pilot sequences to the base station for channel estimation, and \gls{csi} remains constant during a time slot.
        The \gls{mumimo} scheduler then obtains the full knowledge of \gls{csi} of $L$ associated users.
        Using the full \gls{csi} matrix, base station selects a set of $N \le \text{min}(\bar{N}, L)$ among $L$ users to transmit data (either uplink or downlink), where $\bar{N}$ is the maximum number of users that can be selected in one resource block and one time-slot.

        When considering an uplink transmission, base station will receive:
        \begin{equation}
            \mathbf{y} = \mathbf{H} \mathbf{u} + \mathbf{n},
        \end{equation}
        where $\mathbf{u} \in \mathcal{C}^{N}$ are transmitted symbols,
        $\mathbf{H} \in \mathcal{C}^{M \times L}$ is the \gls{csi} matrix,
        $\mathbf{n} \sim C\mathcal{N}(0, \sigma^2 I)$ is an additive white Gaussian noise with the variance $\sigma^2$,
        and finally, $\mathbf{y} \in \mathcal{C}^M$ is the received signal vector at the base station.
        \john{What is M?}
        \thanh{I explained in the previous paragraph. M is number of antennas. }
        Next, base station calculates the zero forcing beamformer using the estimated \gls{csi} matrix $\hat{\mathbf{H}}$ as:
        \begin{equation}
            \mathbf{W} = \hat{\mathbf{H}}(\hat{\mathbf{H}}^H\hat{\mathbf{H}})^{-1}.
        \end{equation}
        After receiving the signal $\mathbf{y}$, base station decodes the transmitted symbols $\mathbf{u}$ via:
        \begin{equation}
            \hat{\mathbf{u}} = \mathbf{W}^H\mathbf{y}.
        \end{equation}
        Then, instantaneous transmission rate achieved by each user $l \in \{1, \ldots, L\}$ in resource block $b \in \{1, \ldots, B\}$ and time slot $t \in \{1, \ldots, T\}$ is calculated as:
        \begin{equation}
            r_{l, b}^t = \text{log}(1 + \text{SINR}_{l, b}^t),
        \end{equation}
        where $\text{SINR}_{l, b}^t$ is the received \gls{sinr} at base station from each selected user $l$ at resource block $b$ and time slot $t$.

    \subsection{MU-MIMO Scheduling Problem}
        Base station needs to select $N$ users among total $L$ users to maximize the throughput while ensuring fairness.
        Let $r_l^t$ be the instantaneous rate user $l$ scheduled with other users.
        Also, let $R_l^t$ represent the average transmission rate that has been allocated to user $l$ up to time slot $t$, then we have:
        \begin{equation}
          R_l^{t+1} = (1 - \beta) r_l^t + \beta R_l^t
            \label{eq:preliminary:average_rate}
        \end{equation}
        \kei{The indicator can be omitted in this formula, since when an user is not scheduled, it $r^t=0$}
        where $x_l^t$ is the binary decision variable representing whether user $l$ is selected for transmission time slot $t$, and $\beta \in [0, 1]$ is the weight to control the importance of instantaneous rate in comparison to the average allocated transmission rate.
        At each time slot $t$, base station solves the following \gls{optpf} problem as follows:
        \begin{equation}
            \begin{aligned}
                \underset{x_l^t, \; \forall l \in \{1, \ldots, L\}}{\text{maximize}} & \quad \sum_{l=1}^{L} \frac{r_l^t x_l^t}{R_l^t},  \\
                \text{subject to} & \quad \sum_{l=1}^{L} x_l^t \leq \bar{N},
            \end{aligned}
            \label{eq:preliminary:problem}
        \end{equation}
        \kei{$\overline{N}$ is not explained. }
        \thanh{It has been explained before Eq. (1). }
        where $x_l^t \in \{0, 1\}$.
        This scheduling problem can be reformulated as an \gls{ilp} problem, which is NP-hard \cite{lau2005proportional}.
        To solve this NP-hard combinatorial optimization problem, previous works trained \gls{drl} agents to infer the near-optimal solution from the problem statement in real-time \cite{an2023deep}.

    \subsection{DRL-based MU-MIMO Scheduler}
        \label{subsec:preliminary:drl}

        To solve this difficult problem in real-time, we consider a recent open-source \gls{drl}-based beam allocation algorithm \cite{an2023deep} using \gls{sac} and \gls{knn}.
        First, this method reformulates the combinatorial optimization problem into the following \gls{mdp}:

        \subsubsection{Observation Space}
            Let $o_l^t := [\gamma_l^t, R_l^t, h_l^t]$ be the observation for each user $l, \forall l \in \{1, \ldots, L\}$, where $\gamma_l^t=\frac{r_l^t}{\bar{r}_l^t}$ is the normalized instantaneous rate, and $\bar{r}_l^t$ is the maximum rate of user $l$ when only $l$ transmits data. $R_l^t$ is the average allocated transmission rate till time $t$ for user $l$, and $h_l^t$ is the \gls{csi} of user $l$ at time $t$.

        \subsubsection{Action Space}
            The action $a^t$ is an integer, which indicates the index to the set of users.
            It takes the value from the action set  $\mathcal{A}=\{1, 2, \ldots, \sum_{i=1}^{\bar{N}} \binom{L}{i}\}$.
            We can decode $a^t$ as a binary variable $x_l^t$, which contains the decision whether the base station select user $l$ to transmit at time slot $t$ or not.
            \john{Not clear, rewrite.}
            \thanh{TODO. }

        \subsubsection{Action Generation}
            \label{subsubsec:action_generation}
            The policy network $\pi(.|o^t; \theta_\pi)$ takes the observation $o^t$ as the input to generate the action $a^t$.
            To design the policy network, a naive approach is to design the policy network with an output size of $|\mathcal{A}|$, where $a^t$ represents the index of the maximum output unit of the policy network.
            However, this approach becomes impractical as the size of the search space $\mathcal{A}$ grows.
            In \cite{an2023deep}, the Wolpertinger architecture \cite{dulac2015deep} is implemented to select actions from a large discrete action space.
            The policy network $\pi(.|o^t; \theta_\pi)$ maps observations to a ``proto action'' $u^t \in [0, 1]^D$ rather than directly mapping the observation space to the large discrete action space, where $D$ is the number of dimension for proto action.
            When $D$ increases, the number of discrete actions per dimension of $u^t$ decreases, which improves the precision of the mapping from continuous proto action $u^t$ to discrete action $a^t$.
            \john{What is D?}
            \thanh{My attempt to fix is above.}
            This proto action serves as a continuous representation of large discrete action space $\mathcal{A}$, requiring only $D$ continuous outputs from the policy network instead of $|\mathcal{A}|$ output units.
            Discrete actions $a^t$ are then placed in the continuous proto-action space in an equal distance.
            For example, if there are $\sum_{i=1}^{\bar{N}} \binom{L}{i}$ discrete actions and the proto action space has $D$ dimensions, then each dimension of the proto action space will be divided in to $\sqrt[D]{\sum_{i=1}^{\bar{N}} \binom{L}{i}}$ equidistant points.
            Then, each combination of points in all $D$ dimensions represents a discrete action.
            The policy network $\pi(.|o^t; \theta_\pi)$ takes the observation $o^t$ as an input and outputs $\hat{u}^t$ as the predicted proto action.
            The corresponding discrete action is one among $k$-nearest neighbor of $\hat{u}^t$ in the proto action space.

        \subsubsection{Action Refinement}
            \label{subsubsec:action_refinement}
            The critic network $Q(o^t, u^t; \theta_Q)$ chooses the best among the $k$ nearest discrete actions from the generated proto action.
            For example, the nearest proto action from the inferred proto action may have a low $Q$-value, while others in the set of $k$-nearest neighbors may all give high $Q$-values.
            To avoid picking outlier actions, the critic network $Q(o^t, u^t; \theta_Q)$ is used to evaluate the $Q$-value of $k$-nearest neighbors from the inferred proto action $\hat{u}^t$.
            The critic network in this Wolpertinger architecture takes the observation and the proto action as the input and produces the estimated $Q$-value.
            Finally, the discrete action that maximizes the $Q$-value among the $k$-nearest neighbors will be chosen.
            Note that \gls{sac} is the training algorithm for actor and critic in this architecture \cite{haarnoja2018soft}.

        \subsubsection{Reward}
            The reward for the centralized \gls{drl}-based scheduler at the base station is defined for each time slot $t$:

            \begin{equation}
                r^t = \sum_{l=1}^{L} \frac{r_l^t x_l^t}{R_l^t} ,
            \end{equation}
            $r_l^t$ is the instantaneous rate of the $l^{\text{th}}$ user time slot $t$, $R_l^t$ is its cumulative amount of transmitted data at $t$.
            By training via this reward function, the \gls{drl}-based scheduler to choose users to maximize the instantaneous rate for users having low cumulative rates, which also provides quality solutions for Problem \ref{eq:preliminary:problem}.

    \subsection{Formal Verification of DNN - A Primal}
        \gls{dnn}, just like traditional software, can have ``bugs'', e.g., producing unexpected results on inputs that are different from those in training data, or small perturbations to the inputs can result in misclassification.
        To address this question, researchers have developed various algorithmic techniques and supporting tools to verify properties of DNNs~\cite{katz2022reluplex,singh2019abstract,duong2023dpll,duong2024harnessing,duong2025neuralsat,duong2025neuralsat2,duong2025compositional,duong2026verifying}.

        There are three main approaches to solve the \gls{dnn} verification problem: (1) constraint-based approaches, (2) abstraction-based approaches, and (3) hybrid approaches.
        Obtaining the exact solution for constraint-based approaches can be time consuming when the size of the given \gls{dnn} increases \cite{katz2022reluplex}.
        Alternatively, abstraction-based techniques and tools use abstract domains, such as polytope (e.g., DeepPoly \cite{singh2019abstract}), to scale up verification while slightly reduce the accuracy.
        Abstract domains such as interval, zonotope, or polytope over-approximate output values of nonlinear units of \gls{dnn}s.
        The main objective of the abstraction is to construct polyhedral over-approximations defined by linear inequalities that tightly enclose the possible outputs of the non-linear functions, e.g., \gls{relu}.
        NeuralSAT \cite{duong2023dpll,duong2024harnessing} combined constraint-based verification and abstraction-based techniques to handle more challenging properties and further increase the size of verifiable \gls{dnn}.

        In addition, abstraction-based verifiers are integrated in supervised training of \gls{dnn} models, such that the result models is easier to verified it robustness properties \cite{xu2024training}.
        However, \gls{dnn} verification have not been applied widely to aid the attack the \gls{drl}-based system.
        To the best of our knowledge, we are the first to leverage \gls{dnn} verification to generate adversarial observations for \gls{drl}.

    \section{Threat Model}
    \label{sec:threat_model}
    This section envisions potential threats for the considered \gls{drl}-based \gls{mumimo} scheduling system (see~\autoref{fig:threat_model:summary}).
    We will explain realistic assumptions on the capabilities and limitations of the adversarial users.

\subsection{The Interactions of Adversaries}
    \begin{figure}[ht]
        \centering
        \includegraphics[width=\linewidth]{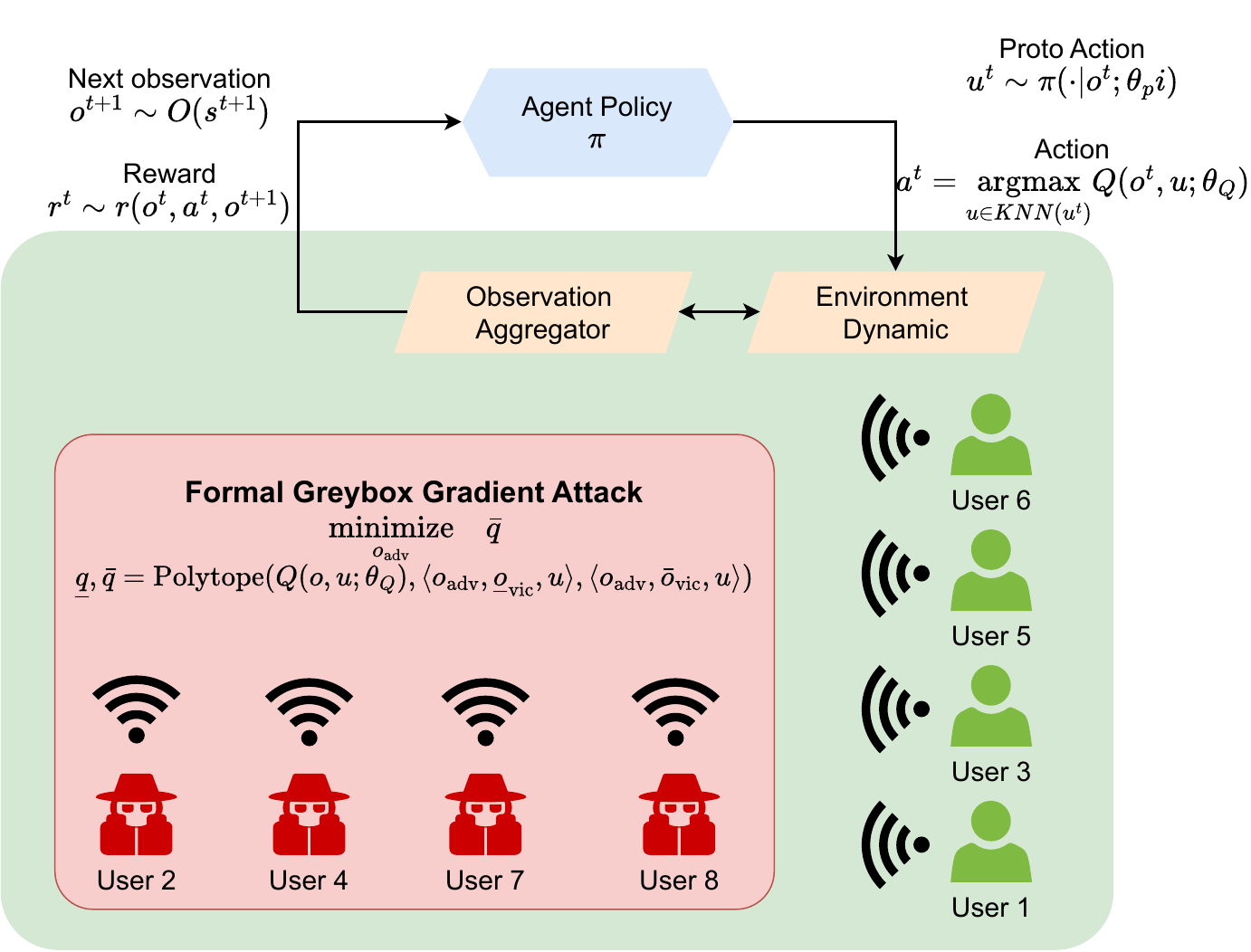}
        \caption{Threat model of \gls{drl}-based scheduler for \gls{mumimo} networks.}
        \label{fig:threat_model:summary}
    \end{figure}
    In our proposed grey-box threat model (\autoref{fig:threat_model:summary}), there exists $L_{adv}$ adversarial users and $L_{vic}$ victim such that $L = L_{\text{adv}} + L_{\text{vic}}$.
    A designated cluster (i.e, one among the attackers) head controls all adversarial users.
    This cluster head obtains the model of a \gls{drl} agent to compute adversarial \gls{csi} values for users under its control so to disrupt network performance for the victims.
    To facilitate the computation of adversarial observations, the cluster head must possess the sufficient computational capability, or it can offload this task to a cloud server.
    Once the adversarial \gls{csi} values are calculated, the cluster head distributes this information to $L_{adv}-1$ other adversarial users.
    Subsequently, the adversarial users are tasked with either reporting the falsified \gls{csi} to the base station or adjusting the \gls{csi} reference signal such that the base station receives the adversarial \gls{csi} calculated by the cluster head.
    Note that this attack strategy provides a single adversarial input at the initiation of the attack, without the need of recalculating during the attack.
    As \gls{fggm} computes the adversarial \gls{csi} for potential victim \gls{csi}s within their normal ranges, the attack remains effective even if the victim \gls{csi}s change due to movement.
    \kei{How often this can be done? Can it be offloaded to the cloud? How about the resource for cluster to distribute CSIs?}
    \thanh{This can be done once. Yes it can be offloaded to the cloud. I have written about it above. As the cluster head only transmit the information once, the resource for distribute CSIs is negligible. }
    \kei{How about CSI changes as user moves?}
    \thanh{I have answered above. }

\subsection{Victim's Observation}
    To search for an adversarial input, attackers need to estimate the \gls{csi} values of target users.
    Prior works have primarily focused on scenarios where attackers know the exact \gls{csi} values of victims \cite{hou2022muster,manoj2021adversarial}.
    This line of work typically involves downlink data transmission, where the base station broadcasts a \gls{csi} reference signal to users.
    users then estimate the \gls{csi} and provide feedback to the base station in plain text, which can potentially be intercepted by adversaries.
    One approach to mitigate this drawback is to encrypt the \gls{csi} feedback at the physical layer to maintain confidentiality from malicious users.
    Obtaining the exact values of victim \gls{csi} is more challenging in the context of uplink transmission in LTE/5G mobile systems, where users send the \gls{csi} reference signal to the base station.
    \john{Explain why it is challenging?}
    \thanh{I explained it below.}
    Since attackers are located in different locations than the base station, the \gls{csi} reference signals they measure will differ from the \gls{csi} measured at the location of the base station.

    In this work, our method estimates the boundaries of victim observations by leveraging the observation normalizer, which is an essential component of a \gls{drl} agent, aiding in stabilizing the training process \cite{mania2018simple}.
    Given that \gls{drl} observations often consist of multiple dimensions, each dimension bounded by an interval, the observation normalizer transforms each dimension to follow a standard Normal distribution.
    This normalization process is similar to data whitening in regression tasks to ensure that all components of the observations have an equal influence.
    Without the observation normalizer, training \gls{drl} algorithms for most continuous control tasks becomes impractical.
    The observation normalizer contains the mean and standard deviation of observations obtained during the training phase.
    Hence, these statistics can be leveraged to estimate the average observation values of a typical user.
    Let $\mu_o$ and $\sigma_o$ denote the mean and standard deviation of the observation vector for a user, which can be extracted from the observation normalizer of the trained \gls{sac} agent.
    We can bound the victim's observations between $\mu_o \pm \delta_o \sigma_o$, where $\delta_o$ is a scaling factor controlling the width of the bounds.
    Similarly, the search space for potential attackers can be defined within the same limits.

    \section{Formal Grey-box Gradient Method}
    \label{sec:propose_attack}
    This section explain how to attack the \gls{drl}-based scheduler for \gls{mumimo} without knowing the exact values of observations of victims.
    We begin by explaining our abstraction-based method to over-approximate the output of a \gls{dnn}.
    Then, we formulate the network throughput degradation attack into a non-linear optimization problem, and solve it via stochastic gradient descent.

    \subsection{Compute DNN Output Bounds via Polytope}

        We opt for the polytope domain (e.g., DeepPoly \cite{singh2019abstract}) as it is a precise and computationally efficient domain for general \gls{dnn} verification problems.
        Note that the notation utilized in this section is distinct from that used in other sections and is applicable solely within the context of this particular section.
        The polytope abstract domain works by:
        (a) representing the input ranges of the \gls{dnn} as polytopes,
        (b) applying transformation rules to the affine and non-linear units of \gls{dnn} to obtain polytope regions representing output values of those units, and finally
        (c) converting the polytope results into output's lower and upper bounds.
        The resulting outputs are over-approximations of the actual outputs.
        This ensures the approximation remains conservative, meaning that the \emph{actual outputs} are always contained within the predicted bounds.
        
        Specifically, the abstraction \emph{computes linear upper and lower bounds of the output of neural network $N$ w.r.t input $x \in \mathcal{X}$}, where $\mathcal{X}$ is bounded input domain:
        \begin{equation}
            \underline{\textbf{A}} x + \underline{\textbf{b}} \le N(x) \le \overline{\textbf{A}} x + \overline{\textbf{b}}, \quad \forall x \in \mathcal{X},
        \end{equation}
        where the output of neural network $N(x)$ is bounded by linear function of input $x$. $\underline{\textbf{A}},\underline{\textbf{b}}$ denote the coefficients for the lower bound hyper-plane, whereas $\overline{\textbf{b}}$. $\underline{\textbf{A}}$ are those of the upper bound hyper-plane. 
        
        To get those of an $L$-layer feed-forward neural network $N$, it propagates bounds of $N(x)$ as a linear function to the output of each layer, in a \emph{backward} manner.
        Let $h^{(i)}_j$ be the pre-activation of the $j$-th ReLU neuron in $i$-th layer, and $g^{(i)}_j$ to denote the post-activation values.
        To compute the bounds for each layer, we start at the output layer $N(x)$ (or $h^{(L)}(x)$) with the following identity linear relationship:
        \begin{equation}
             h^{(L)}(x) \le N(x) \le  h^{(L)}(x), \quad \forall x \in \mathcal{X}.
        \end{equation}
        
        The next step then is to backward propagate from the identity linear relationship through the linear layer $h^{(L)}(x) = \textbf{W}^{(L)} g^{(L-1)}(x)$ to get the linear bounds of $N(x)$ w.r.t $g^{(L-1)}(x)$ (for simplicity, we ignore biases):
        \begin{equation}
            \textbf{W}^{(L)} g^{(L-1)}(x) \le N(x) \le \textbf{W}^{(L)} g^{(L-1)}(x), \quad \forall x \in \mathcal{X}.
        \end{equation}

        To get the linear bounds of $N(x)$ w.r.t $h^{(L-1)}(x)$, we need to backward propagate through ReLU layer $g^{(L-1)}(x) = ReLU(h^{(L-1)}(x))$.
        Since it is nonlinear, we perform an approximation.
        Suppose that $\textbf{l}^{(L-1)} \le h^{(L-1)}(x) \le \mathscr{u}^{(L-1)}$ are intermediate pre-activation bounds of $h^{(L-1)}(x)$.
        If $h^{(L-1)}(x)$ can only have positive ($\textbf{l}^{(L-1)} > 0$) or non-positive values ($\textbf{u}^{(L-1)} \le 0$), then $g^{(L-1)}(x) = h^{(L-1)}(x)$ or $g^{(L-1)}(x) = 0$, respectively.

        When $h^{(L-1)}(x)$ can have both positive and negative values ($\textbf{l}^{(L-1)} < 0 < \textbf{u}^{(L-1)}$), we have:
        \begin{equation} \label{eq:bounds}
            \small
            \begin{aligned}
                &g^{(L-1)}(x) \ge {\alpha^{(L-1)}} h^{(L-1)}(x), \\
                &g^{(L-1)}(x) \le \frac{\textbf{u}^{(L-1)}}{\textbf{u}^{(L-1)} - \textbf{l}^{(L-1)}} h^{(L-1)}(x) - \frac{\textbf{u}^{(L-1)} \textbf{l}^{(L-1)}}{ \textbf{u}^{(L-1)} - \textbf{l}^{(L-1)}},
            \end{aligned}
        \end{equation}
        where the slope ${\alpha^{(L-1)}}$ is a vector of 0 or 1 as defined in~\cite{singh2019abstract}.
        With these linear relaxations, we can get the linear equation of output $N(x)$ w.r.t $h^{(L-1)}(x)$:
        \begin{equation}
            \begin{aligned}
                &N(x) \ge \textbf{W}^{(L)} \underline{\textbf{D}}_{\alpha}^{(L-1)} h^{(L-1)}(x) + \underline{\textbf{b}}^{(L-1)}, \\
                &N(x) \le \textbf{W}^{(L)} \overline{\textbf{D}}_{\alpha}^{(L-1)} h^{(L-1)}(x) + \overline{\textbf{b}}^{(L-1)},
            \end{aligned}
        \end{equation}
        where
        \begin{equation}
            \underline{\textbf{D}}_{\alpha, (j, j)}^{(L-1)} = \begin{cases}
                {\alpha^{(L-1)}_j}                                           , & \textbf{W}^{(L)}_j \ge 0, \\
                \frac{\textbf{u}^{(L-1)}_j}{\textbf{u}^{(L-1)}_j - \textbf{l}^{(L-1)}_j}    , & \textbf{W}^{(L)}_j < 0,
            \end{cases}
        \end{equation}
        \begin{equation}
            \underline{\textbf{b}}^{(L-1)} = \underline{\textbf{b}}^{\prime (L-1) \top} \underline{\textbf{W}}^{(L)},
        \end{equation}
        \begin{equation}
            \underline{\textbf{b}}^{\prime (L-1)}_j = \begin{cases}
                0, & \textbf{W}^{(L)}_j \ge 0, \\
                - \frac{\textbf{u}^{(L-1)}_j \textbf{l}^{(L-1)}_j}{ \textbf{u}^{(L-1)}_j - \textbf{l}^{(L-1)}_j}, & \textbf{W}^{(L)}_j < 0.
            \end{cases}
        \end{equation}
        The diagonal matrices $ \underline{\textbf{D}}_{\alpha}^{(L-1)},  \overline{\textbf{D}}_{\alpha}^{(L-1)}$ and biases $\underline{\textbf{b}}^{(L-1)}, \overline{\textbf{b}}^{(L-1)}$ reflect the linear relaxations and also consider the signs in $\textbf{W}^{(L)}$ to maintain the lower and upper bounds.
        The definitions for $j$-th diagonal element of $\overline{\textbf{D}}_{\alpha, (j, j)}^{(L-1)}$ and bias $\overline{\textbf{b}}^{(L-1)}$ are similar, with the conditions for checking the signs of $\textbf{W}^{(L)}_j$ swapped.

        The abstraction to backward propagate these lower and upper bounds throughout the network, layer by layer (e.g., $h^{(L-2)}(x)$, $g^{(L-2)}(x)$, etc.), until reaching the input $g^{(0)}(x) = x$, getting the eventual bounding linear equations ($\underline{\textbf{A}}x + \underline{\textbf{b}}$ and $\overline{\textbf{A}}x + \overline{\textbf{b}}$) of $N(x)$ in terms of input $x$.

    \subsection{Optimization Problem to Degrade Victim Throughput}
        \label{subsec:propose:problem}
        The adversaries adjust their observations to cripple the throughput of the victims by minimizing the probability that victims are selected for data transmission, meaning the $Q$-value of any action $a_t \in \mathcal{A}_{\text{vic}}$ containing a victim.
        Specifically, let $o_{\text{adv}}$ represent the local observations of the adversaries and let $\underline{o}_{\text{vic}}$ and $\overline{o}_{\text{vic}}$ denote the lower and upper bounds of the local observations of the victims, respectively.
        With these inputs, the adversarial cluster head computes the upper and lower bounds of the continuous proto actions ($\underline{u}$ and $\overline{u}$) using polytopes over-approximation as follows:
        \begin{equation}
            \small
            \underline{u}, \overline{u} = \textsc{Polytope}(\pi(.|o; \theta_\pi),
                                         \langle o_{\text{adv}}, \underline{o}_{\text{vic}} \rangle ,
                                         \langle o_{\text{adv}}, \overline{o}_{\text{vic}}\rangle ),
            \label{eq:propose:dp_u}
        \end{equation}
        where $\pi(.|o; \theta_\pi)$ represents the trained actor models, $\langle o_{\text{adv}}, \underline{o}_{\text{vic}} \rangle$ and $\langle o_{\text{adv}}, \overline{o}_{\text{vic}} \rangle$ are the concatenations of adversarial inputs with the intervals of victim inputs.
        By leveraging the trained critic network $Q(o, u; \theta_Q)$, \gls{fggm} assesses the $Q$-value based on the inputs and proto actions.
        The final optimization problem for generating adversarial observations is as follows:
        \begin{equation}
            \small
            \underset{o_{\text{adv}}}{\text{minimize}} \quad \overline{q},
        \end{equation}
        such that
        \begin{equation}
            \small
                \underline{q}, \overline{q} = \textsc{Polytope}(Q(o, u; \theta_Q),
                    \langle o_{\text{adv}}, \underline{o}_{\text{vic}}, u_i \rangle ,
                    \langle o_{\text{adv}}, \overline{o}_{\text{vic}}, u_i \rangle ),
        \end{equation}
        and
        \begin{equation}
            \small
            \mu_{\text{adv}}-\delta_{\text{adv}}\sigma_{\text{adv}} \le o_{\text{adv}} \le \mu_{\text{adv}}+\delta_{\text{adv}}\sigma_{\text{adv}},
        \end{equation}
        for all
        \begin{equation}
            \small
            \begin{cases}
                \underline{u} & \le u_i \le \overline{u}, \\
                a_i & \in \mathcal{A}_{\text{vic}},
            \end{cases}
        \end{equation}
        where $u_i$ is the proto action that corresponds to the discrete action $a_i$, $\mathcal{A}_{\text{vic}} \subset \mathcal{A}$ is the subset of actions that includes all actions containing victims, and $[\underline{q}, \overline{q}]$ in the estimated interval of $Q$-values.
        The values of the observation for adversaries should be clipped to fall within the desired range $[\mu_{\text{adv}}-\delta_{\text{adv}}\sigma_{\text{adv}}, \mu_{\text{adv}}+\delta_{\text{adv}}\sigma_{\text{adv}}]$, where $\mu_{\text{adv}}$ and $\sigma_{\text{adv}}$ are the mean and standard deviation of adversarial observations, and $\delta_{\text{adv}}$ is a hyper-parameter to control the width of the search space.
        The objective is to minimize the upper bound of the $Q$-value for all possible actions containing victims.
        
        Given that $o_{\text{adv}}$ is differentiable, we can solve this problem using a standard gradient-based optimizer \cite{kingma2014adam}. 
        Since this optimization problem is non-linear, \gls{fggm} restarts many times to optimize $o_{\text{adv}}$ from multiple initial values. 
        Note that polytope abstraction ensures that the upper bound is always higher than the lower bound, minimizing only the upper bound $Q$-value leads to the reduction of the lower bound, and it is sufficient for the attack.
        Furthermore, the cluster head only needs to solve this problem once at the beginning of the attack period and the group of adversarial users can reuse the optimized adversarial observations for all time steps.

    \section{Evaluation and Discussion}
\label{sec:experiment}
We evaluate our adversarial attacks \gls{drl}-based \gls{mumimo} scheduler, and show that the proposed attack is effective in reducing network access probability and degrading throughput of victims.
We focus our evaluation on the following research questions:

\textbf{RQ1} (\autoref{subsec:rq1}): How does \gls{fggm} perform under various sizes of the search space for adversarial observations, and various sizes of bounds for observations of victims?

\textbf{RQ2} (\autoref{subsec:rq2}): Can throughput degradation attack be performed more effectively if there are more adversarial users using \gls{fggm}?

\textbf{RQ3} (\autoref{subsec:rq3}): What is the most significant consequence if the \gls{drl}-based scheduler is operating with 50\% \gls{fggm} attackers?

\subsection{Experimental Design}

    \subsubsection{Benchmarking Environment}
        To access our proposed attack, we created the target system based on a publicly available \gls{drl} algorithm for large-scale discrete action space \cite{dulac2015deep}, and trained this algorithm for the \gls{mumimo} scheduling problem \cite{an2023deep}.
        The \gls{csi} values of all users were simulated using \gls{quadriga} software \cite{jaeckel2014quadriga}.
        We specifically examined a scenario where mobile users move at an average speed of $2.8$ m/s with random directions.
        In this scenario, mobile users rebounded upon reaching the boundary, which is a circular area with a radius of 300 meters.
        Additionally, the environment utilized a 16-QAM modulation scheme, and the error vector magnitude of the received constellation is used to derive the signal-to-noise ratio.
        Further details on the hyper-parameters are presented in~\autoref{tab:experiment:parameter}.
        \begin{table}[ht]
            \caption{Simulation parameters}
            \centering
            \begin{tabular}{@{}cc@{}}
                \toprule
                \textbf{Parameter}                           & \textbf{Values} \\ \midrule
                 bandwidth                                   & 20 MHz \\
                 carrier frequency                           & 3.6 MHz \\
                 slot duration                               & 1ms \\
                 number of time slots & 500 \\
                 number of users ($L$)                   & $\{8, 16\}$\\
                 number of antennas ($M$)                    & $\{4, 16\}$\\
                 maximum users per time slot ($\bar{N}$) & $\{4, 4\}$ \\
                 proportional fairness coefficient $\beta$   & $0.5$ \\
                 \midrule
                 number of dimension of proto action $D$     & $\{3, 8\}$ \\
                 number of nearest neighbor $K$              & $\{20, 20\}$ \\
                 number of hidden units in Actor             & $128$ \\
                 number of hidden units in Critic            & $256$ \\
                 number of hidden layers in Actor and Critic & $2$ \\
                 activation function in Actor and Critic     & ReLU \\
                 \midrule
                 number of adversarial users ($L_{\text{adv}}$) & up to $\{4, 8\}$\\
                 number of \gls{fggm}/\texttt{SPGD} iterations& 300 \\
                 number of \gls{fggm}/\texttt{SPGD} restarts & 10 \\
                 number of \texttt{SPGD} samples             & 100 \\
                 \bottomrule
            \end{tabular}
            \label{tab:experiment:parameter}
        \end{table}

    \subsubsection{Baselines}
        The performance of our proposed attack \gls{fggm} is compared with that of conventional schedulers and other potential schemes for generating adversarial observations. Conventional schedulers include policies such as \texttt{Random}, \gls{optpf}, \gls{optmr}, \gls{optpfug}, and \gls{sac}.
        \kei{\gls{optmr} change to optimal capacity over interference (OptCI)}
        \thanh{TODO, as I need to redraw figures.}
        In each time slot $t$, the \texttt{Random} policy randomly selects one action from the action set $\mathcal{A}$.
        \gls{optmr} and \gls{optpf} policies exhaustively evaluate the action set $\mathcal{A}$ to determine the action that yields the highest data rate or proportional fairness score (\autoref{eq:preliminary:problem}).
        However, \gls{optmr} and \gls{optpf} are not scalable for more users as their computational complexities are $\mathcal{O}(2^L)$, where $L$ is the number of total users sharing the same resource block.
        \gls{optpfug} reduces the complexity by grouping users with channel correlation less than $0.5$ in the same group and selecting the group providing the highest proportional fairness score.
        Similarly, \gls{sac} is a scalable method for \gls{mumimo} scheduling, as it takes a polynomial time to infer the set of users from the trained actor and critic networks (\autoref{subsec:preliminary:drl}).

        Three adversarial attack schemes are considered: (1) \texttt{Noise}, (2) \gls{spgd}, and (3) our proposed attack \gls{fggm}.
        \texttt{Noise} generates random \gls{csi} at each time slot $t$ for adversarial users to disrupt the \gls{sac} scheduler, regardless of the observations of the victims, and it represents the class of black-box attacks.
        \gls{spgd} is our adapted version of the origin \gls{pgd} \cite{madry2017towards} to perform grey-box attack.
        \kei{any reference?} \thanh{I added the reference}
        It randomly generates samples of \gls{csi}s within the estimated bounds of the observations of victims and then solves the same problem in \autoref{subsec:propose:problem} for the generated set of samples.
        To ensure fairness in our comparison, \gls{spgd} and \gls{fggm} have the same number of iterations and number of restarts (\autoref{tab:experiment:parameter}).

    \subsubsection{Evaluation metrics}
        We compare the performance of \gls{mumimo} schedulers and attack schemes via four main metrics; namely, proportional fairness, Jain fairness index (JFI), the selection probability of victims, and average data rate of victims.
        To compute these metrics, we run the simulator for 500 time slots using the same \gls{csi} trace from \gls{quadriga} \cite{an2023deep, jaeckel2014quadriga}, and record the set of selected users as well as each of their instantaneous transmission rates.
        The proportional fairness per each time slot is calculated according to \autoref{eq:preliminary:problem}, and the Jain fairness index at each time slot $t$ is defined as:
        \kei{Should it be calculated at the end?}
        \thanh{Yes, it should be. I will change the plot from boxplot of distribution of JFI throughout many time slot to barchart contain the JFI at last step. }
        \begin{equation}
            JFI^t = \frac{(\sum_{l=1}^L R_l^t)^2}{L \sum_{l=1}^L(R_l^t)^2},
        \end{equation}
        where $R_l^t$ is the long-term average rate in \autoref{eq:preliminary:average_rate}.

\subsection{RQ1: Performance under Varying Search Space}
    \label{subsec:rq1}
    \begin{figure}[ht]
        \centering
            \begin{subfigure}[b]{0.6\linewidth}
                \centering
                \includegraphics[width=\linewidth]{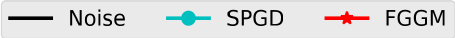}
            \end{subfigure}
            \hfill
            \begin{subfigure}[b]{0.48\linewidth}
                \centering
                \includegraphics[width=\linewidth]{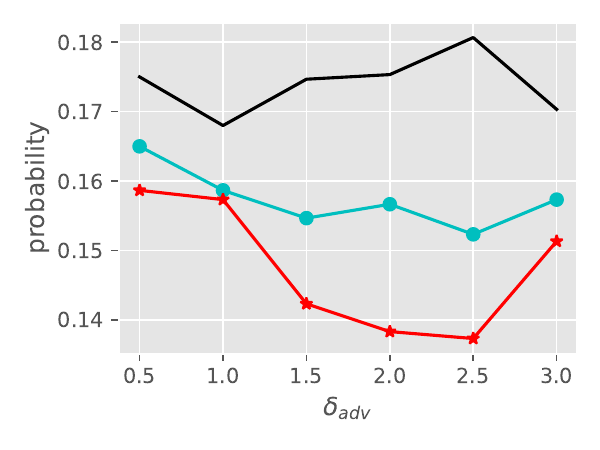}
                \caption{Varying adversary bounds}
                \label{fig:evaluation:rq1:a}
            \end{subfigure}
            \hfill
            \begin{subfigure}[b]{0.48\linewidth}
                \centering
                \includegraphics[width=\linewidth]{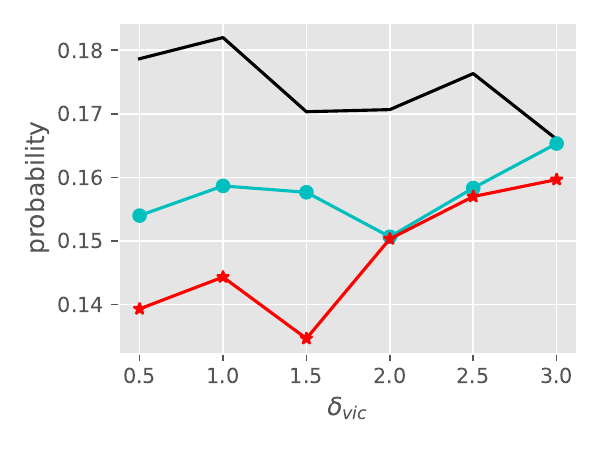}
                \caption{Varying victim bounds}
                \label{fig:evaluation:rq1:b}
            \end{subfigure}
        \caption{The average selection probability of victims when attacker using different attacking schemes to create adversarial CSIs. $L=8, M=4$, and $\bar{N}=4$. $\delta_{vic}$ and $\delta_{adv} \in \{0.5, 1.0, 1.5, 2.0, 2.5, 3.0\}$. Lower is better.}
        \label{fig:evaluation:rq1}
    \end{figure}

    The performances of three attacking schemes are analyzed in \autoref{fig:evaluation:rq1} as the search space sizes vary for observations of adversaries $\delta_{adv}$ and estimate errors for observations of victims $\delta_{vic}$.
    Since $\delta_{adv}$ is measured in terms of standard deviations that adversarial observations can deviate from the average, the searching domain for adversarial users is defined as $\mu_{adv} \pm \delta_{adv} \sigma_{adv}$, where $\mu_{adv}$ and $\sigma_{adv}$ represent the average and standard deviation of the observations of adversarial users, respectively.
    Similarly, $\delta_{vic}$ describes the estimated range within which the observations of victims belong, given by $\mu_{vic} \pm \delta_{vic} \sigma_{vic}$.

    A grid search is conducted on $6^2$ different configurations.
    \autoref{fig:evaluation:rq1:a} illustrates the victim selection probabilities for each value of $\delta_{adv}$ averaged across different values of $\delta_{vic}$.
    Likewise, \autoref{fig:evaluation:rq1:b} demonstrates the victim selection probabilities for each value of $\delta_{vic}$ averaged across different values of $\delta_{adv}$.
    Overall, both figures show that \gls{fggm} facilitated a lower victim selection probabilities than the comparing schemes in all configurations.
    Additionally, both grey-box attacks \gls{spgd} and \gls{fggm} provided strictly better adversarial inputs than \texttt{Noise}.

    We would like to draw readers' attention to the plot for the varying adversarial bounds in \autoref{fig:evaluation:rq1:a}.
    The victim selection probabilities of \gls{fggm} gradually decreased from the starting point, bottomed out at $\delta_{adv}=2.5$ before increasing again towards the end.
    As anticipated, our experiments show that \gls{fggm} would not be able to search for good adversarial observations when the search space was small ($\delta_{adv} \le 1.0$) as there does not exist any satisfied solution.
    \gls{fggm} struggled at $\delta_{adv}=3$ because the stochastic gradient descent algorithm cannot navigate this large search space.
    At the medium search space configuration such as $\delta_{adv}=2.0$, \gls{fggm} achieved a low victim access probability at $13.8\%$, while \gls{spgd} and \texttt{Noise} reached $15.7\%$ and $17.5\%$, respectively.

    Regarding the ablation study for victim bounds in \autoref{fig:evaluation:rq1:b}., \gls{fggm} only exceeded \gls{spgd} with a wide margin when $\delta_{vic} \le 1.5$.
    This is consistent with the literature on \gls{dnn} verification, as wider input bounds lead to huge errors in estimating the output bounds, which reduce the effectiveness of \gls{dnn} verification.

    \begin{tcolorbox}[left=1pt,right=1pt,top=1pt,bottom=1pt]
    \textbf{RQ1 Findings}:
        \gls{fggm} produced a strictly lower victim access probability than \gls{spgd} and \texttt{Noise}.
        Under the best parameter, \gls{fggm} was 21.1\% better than black-box attack and 12.1\% better than \gls{spgd} in terms of victim access probability.
    \end{tcolorbox}

\subsection{RQ2: Performance under Varying Number of Attackers}
    \label{subsec:rq2}
    \begin{figure}[ht]
        \centering
            \begin{subfigure}[b]{\linewidth}
                \centering
                \includegraphics[width=0.6\linewidth]{figure/experiment/low_8_4_legend_ablation_rate.pdf}
            \end{subfigure}
            \hfill
            \begin{subfigure}[b]{0.48\linewidth}
                \centering
                \includegraphics[width=\linewidth]{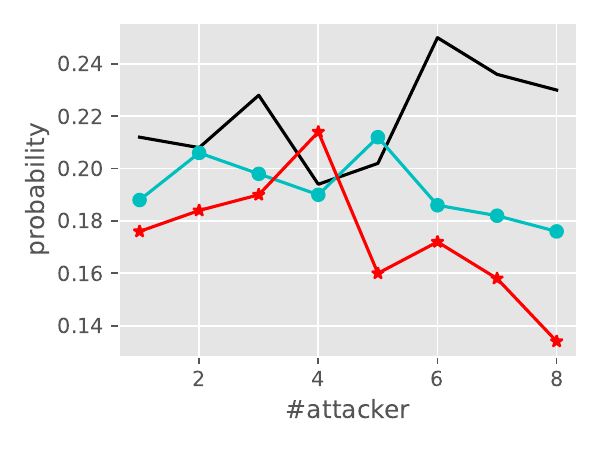}
                \caption{Selection probability}
                \label{fig:evaluation:rq2:a}
            \end{subfigure}
            \hfill
            \begin{subfigure}[b]{0.48\linewidth}
                \centering
                \includegraphics[width=\linewidth]{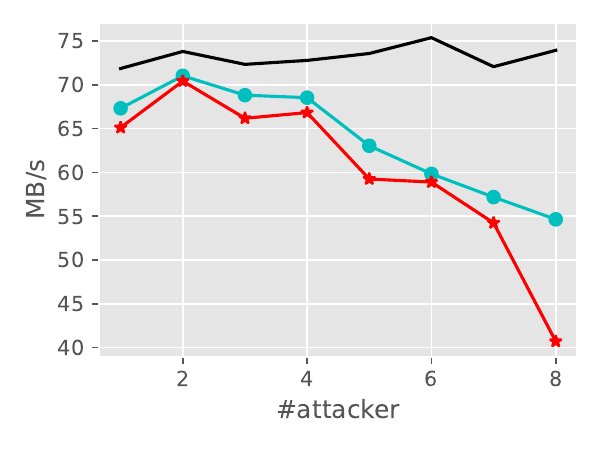}
                \caption{Transmission rate}
                \label{fig:evaluation:rq2:b}
            \end{subfigure}
        \caption{The minimum probability and transmission rate of victims as the number of attackers increased from 1 to 8. $L=16$, $M=16$, and $\bar{N}=4$. }
        \label{fig:evaluation:rq2}
    \end{figure}

    From now on, we select the best parameter for the following experiments: $\delta_{adv}=2.0$ and $\delta_{vic}=1.5$.

    \autoref{fig:evaluation:rq2} presents an ablation study on the impact of the number of attackers on victims' performances.
    In general, both gradient-based grey-box attack methods showed significant improvements compared to \texttt{Noise}, and our proposed \gls{fggm} method surpassed \gls{spgd} in almost all searched numbers of attackers.
    In the presence of a grey-box attack, there is a correlation between the number of attackers and the reduction in selection probabilities and transmission rates of victims.
    In contrast, increasing number of attackers in \texttt{Noise} did not proportionally decrease service quality.
    Note that, even with only one adversarial user, the \gls{fggm} slashed the transmission rate of victims by 9.5\% more than those of \texttt{Noise} scheme.

    In the most favorable scenario where adversarial users occupied half of the considered resource block, the worst transmission rate of victims reduced to half of that of the \texttt{Noise} attacking scheme.
    The rationale behind this trend is that higher number of attackers means smaller number of unknown inputs, leading to a smaller input domain.
    It enables \gls{fggm} to estimate the upper bound of $Q$-value more precisely, further diminishing the throughput and access probability of victims.

\subsection{RQ3: Impact under 50\% Attack}
    \label{subsec:rq3}

    \begin{figure*}[ht]
        \centering
            \begin{subfigure}[b]{0.31\linewidth}
                \centering
                \includegraphics[width=\linewidth]{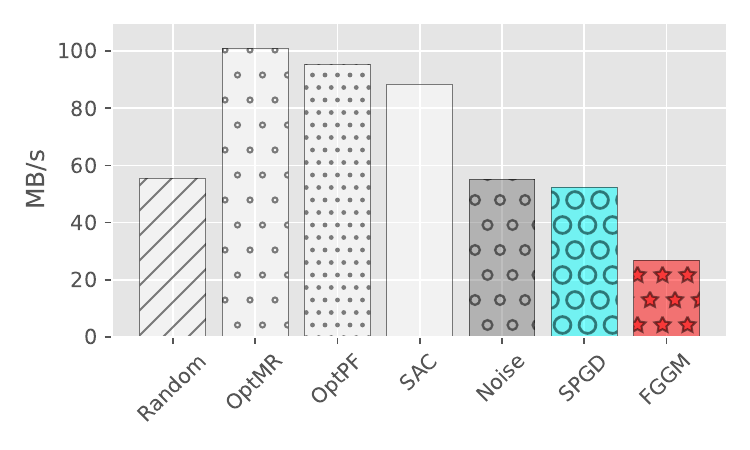}
                \caption{Worst victim delivered data rate}
                \label{fig:8x4:a}
            \end{subfigure}
            \hfill
            \begin{subfigure}[b]{0.31\linewidth}
                \centering
                \includegraphics[width=\linewidth]{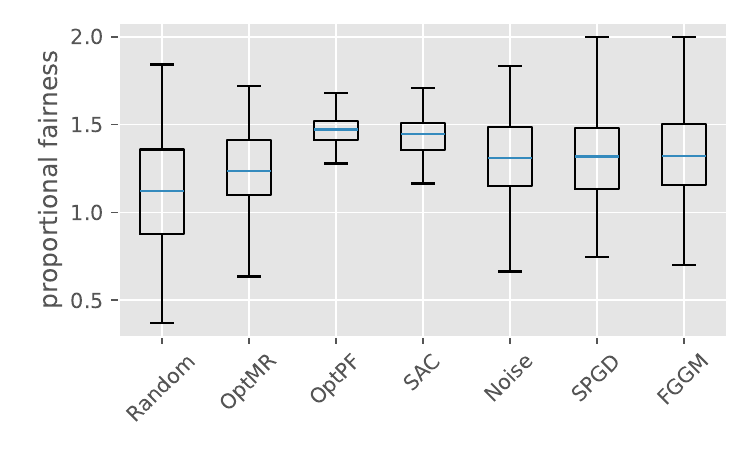}
                \caption{Proportional fairness}
                \label{fig:8x4:b}
            \end{subfigure}
            \hfill
            \begin{subfigure}[b]{0.31\linewidth}
                \centering
                \includegraphics[width=\linewidth]{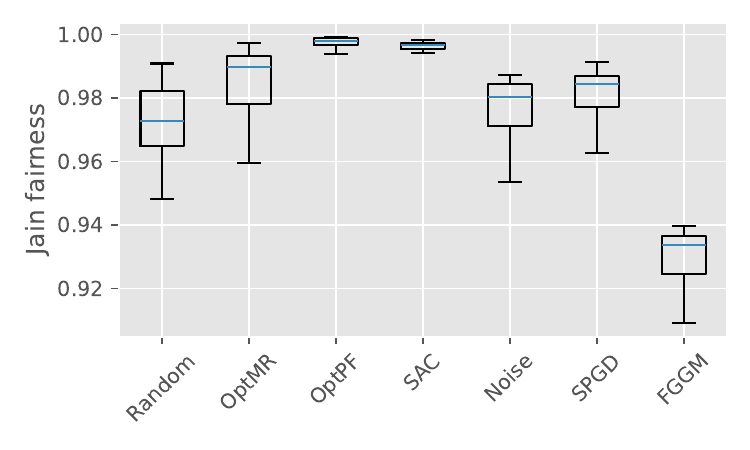}
                \caption{Jain fairness index}
                \label{fig:8x4:c}
            \end{subfigure}
        \caption{Performance of all schedulers and attacking schemes in networks configuration with $L=8$ $M=4$, $\bar{N}=4$, and $L_{adv}=4$}
        \label{fig:8x4}
    \end{figure*}
    At this point, we discuss the impact of our attack to \gls{drl}-based \gls{mumimo} system under the best parameters obtained from the previous ablation studies.
    \autoref{fig:8x4} shows the disruption of the attack in our small setting of 8 users per resource block.
    We observed that without disruption, the \gls{sac} scheduler provided nearly similar performance to \gls{optpf} scheduler in terms of worst victim throughput, proportional fairness, and Jain fairness index.
    This suggests that our \gls{drl}-based scheduler has been well trained.
    However, the \texttt{Noise} attacker degraded the performance of \gls{sac} back to the same level as \texttt{Random}.
    The main reason is that our standard \gls{sac} algorithm for large-scale discrete action problems lacks adversarial training \cite{oikarinen2021robust}, and it also highlights the emergent need for robust \gls{drl} training.
    As shown in \autoref{fig:8x4:a}, \gls{spgd} degrades \gls{sac} by reducing the data rate from 88.5 MB/s to just under 52.3 MB/s, while \gls{fggm}, with the formal verification technique, further halves the throughput of the victim, at an impressive 70\%, which is about 26.6 MB/s.

    \begin{figure*}[ht]
        \centering
            \begin{subfigure}[b]{0.31\linewidth}
                \centering
                \includegraphics[width=\linewidth]{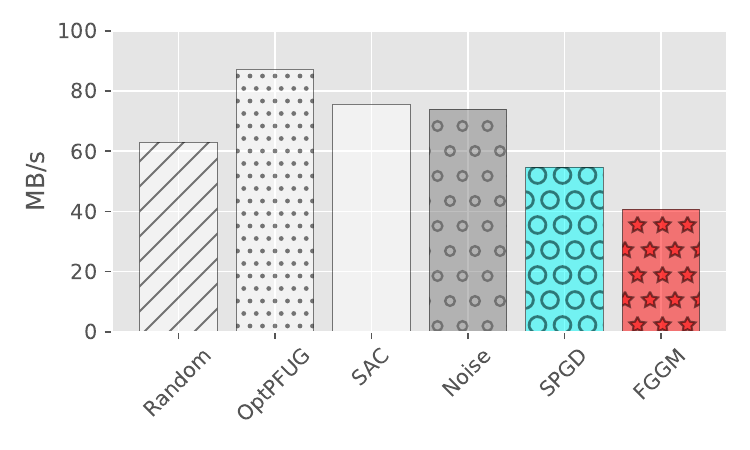}
                \caption{Worst victim delivered data rate}
                \label{fig:16x8:a}
            \end{subfigure}
            \hfill
            \begin{subfigure}[b]{0.31\linewidth}
                \centering
                \includegraphics[width=\linewidth]{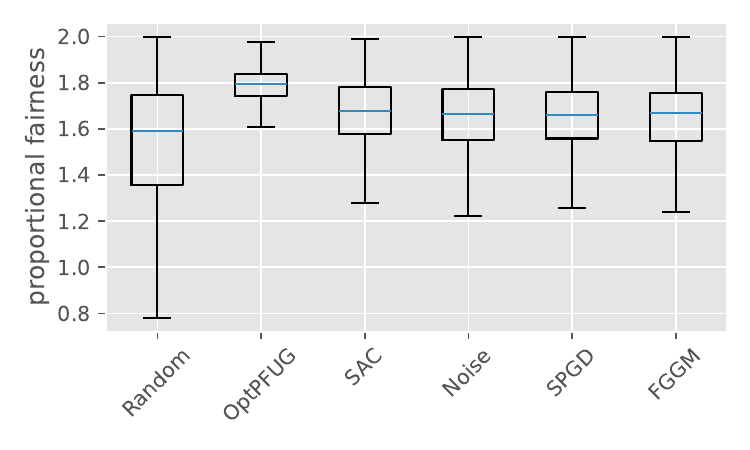}
                \caption{Proportional fairness}
                \label{fig:16x8:b}
            \end{subfigure}
            \hfill
            \begin{subfigure}[b]{0.31\linewidth}
                \centering
                \includegraphics[width=\linewidth]{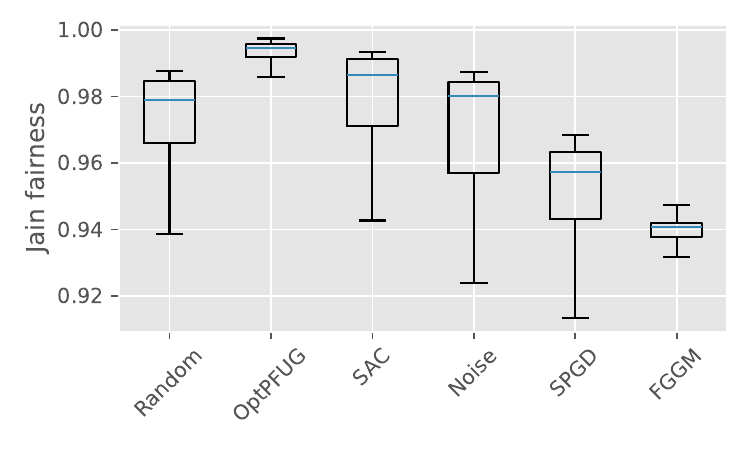}
                \caption{Jain fairness index}
                \label{fig:16x8:c}
            \end{subfigure}
        \caption{Performance of all schedulers and attacking schemes in networks configuration with $L=16$ $M=16$, $\bar{N}=4$, and $L_{adv}=8$}
        \label{fig:16x8}
    \end{figure*}
    In \autoref{fig:16x8}, as the number of users and attackers doubled, the pattern remained the same.
    For instance, the \gls{fggm} significantly reduced the worst victim delivered data rate and Jain fairness index.
    \gls{spgd}  showed improvement from \texttt{Noise}, but \gls{fggm} was still stand out as the best attacking method.
    It demonstrates that minimizing the upper bound of the output Q-value is more effective than sampling a limited number of examples for a \gls{pgd} attack.
    \autoref{fig:16x8:a} indicates that \gls{fggm} still reduced the victim's throughput by 46\% in compared with the original \gls{drl}-based \gls{mumimo} scheduler.
    From the box plots in \autoref{fig:16x8:b} and \autoref{fig:16x8:c}, we observe that \gls{fggm} attack notably reduced Jain fairness distribution of \gls{sac}  but did not have any impact on proportional fairness distribution.
    This implies that training the \gls{drl}-based to maximize multiple fairness metrics may improve the robustness of the \gls{drl} agent against adversarial attacks.

    \begin{tcolorbox}[left=1pt,right=1pt,top=1pt,bottom=1pt]
    \textbf{RQ3 Findings}:
        Under 50\% attack, \gls{fggm} reduce victim data rate from about 46\% to 70\%.
        In resource block with 8 users, data rates of \gls{drl}-based scheduler were crippled from 88.5 MB/s to merely 26.6 MB/s.
    \end{tcolorbox}

    \section{Related works}
    \label{sec:related_work}
    We review several algorithms that are based on \gls{dnn} for solving resource allocation problems in \gls{mumimo}, with the emphasis on their potential unsanitized inputs.
    We then discuss related works on exploiting of these algorithms.

    \subsection{DNN for Resource Allocation in MU-MIMO}
        The multi-user scheduling problem presents a significant challenge in resource allocation due to its computational complexity.
        To reduce the complexity of traditional optimization-oriented methods, many papers advocate \gls{dnn}-based wireless resource allocation algorithms for different problems.
        Recenly, \gls{drl} has been introduced as a solution to approximate the optimal allocation strategy \cite{bello2016neural}.
        For instance, in \cite{shi2018machine}, the Levenberg-Marquardt algorithm is utilized to optimize a fully connected neural network that maps the channel state information (\gls{csi}) to the user set to maximize the sum rate.
        Alternatively, \cite{bu2019reinforcement} proposes a method where users directly report their \gls{csi} to the base station (base station), which trains a \gls{drl} model to select a subset of users for transmission to maximize the sum rate.
        Moreover, in \cite{guo2020novel}, a real-time \gls{ddpg} model is developed with \gls{csi} as the main input for user selection during transmission.
        Similarly, \cite{chen2021deep} introduces a \gls{sac} network that utilized \gls{csi} for user selection.
        Conversely, in \cite{chen2021joint}, previous achievable rates and channel correlation matrices derived from \gls{csi} is used to determine both the user set and their channel precoding strategy.
        Additionally, \cite{huang2023drl} incorporates channel quality index, data amount, and data type as the state variables in the selection algorithm across multiple layers.
        Lastly, \cite{an2023deep} implements a vanilla SMART approach where \gls{csi} served as input for a \gls{dnn} actor, showcasing one of the most effective methods for multi-user scheduling.

        In summary, the input of recently proposes \gls{dnn}-based resource allocation algorithms in massive \gls{mimo} included the \gls{csi}, large scale fading coefficient, spectral noise, \gls{sinr}, or transmission rate.
        Note that they can be modified by a group of adversarial users, and so the \gls{dnn} output could be manipulated to achieve malicious goals.
        The next section provides an overview of existing attacks to the input of \gls{dnn}-based resource allocation algorithms.%

\subsection{Attacking Algorithms for Resource Allocation in MU-MIMO}
        Recent works proposed attacks on \gls{dnn}-based wireless resource allocation algorithms such as power allocation, resource slicing, and multi-user scheduling.

        In \cite{manoj2021adversarial}, the authors conduct a white-box attack on a \gls{dnn}-based regression model that learned a mapping from users' geological positions to the optimal power allocation, obtained by solving the problem using traditional optimization tools.
        \kei{Is there any comparison with this white-box attack?}
        \thanh{Yes, I compared with PGD, but it is a sampling-based version of PGD to fit our problem and threat model.}
        The threat model involves attackers overwriting the positioning information of nearby users so to manipulate the network input using \gls{fgsm} \cite{huang2017adversarial,goodfellow2014explaining}, \gls{pgd} \cite{madry2017towards}, and \gls{mifgsm} \cite{dong2018boosting}.
        Similarly, other works \cite{liu2020adversarial, kim2021adversarial} proposes gradient-based white-box attacks on the full input of the \gls{dnn}, while maintaining the assumption of knowing the exact location of all users, which is impractical.
        To address this issue, \cite{santos2021universal} proposes a black-box attack based on \gls{uap} \cite{moosavi2017universal}.
        However, the evaluation reveals that the black-box attack lacks of stealthiness as it is effective only with large perturbation magnitudes.
        Additionally, previous works allow the adversary to apply adversarial perturbations to the full input of the \gls{dnn}.

        In \cite{shi2022attack}, a resource slicing attack based on Q-learning was carried out.
        To reduce the Q-learning resource slicing agent's reward, they suggests to train another Q-learning agent as an attacker.
        This Q-learning-based attacker selects the resource block and sent a jamming signal within a limited energy budget.
        Subsequently, \cite{wang2023robust} introduces a \gls{madrl} resource slicing system, where each agent represented a base station in a multi-cell architecture.
        These agents are trained using centralized training decentralized execution to map user requests to available channel resources, maximizing the system's throughput.
        An attacker is also trained to select resources to jam in order to minimize the system's global reward.
        Note that training an attacker is computationally intensive and may not effectively exploit vulnerabilities in trained resource slicing agents.

        In \cite{hou2022muster}, a black-box heuristic attack is proposed to target the heuristic selection of users in multi-user scheduling in massive \gls{mimo} networks.
        The advantage of this work is that only a few attackers need to manipulate the \gls{csi} in the networks.
        However, they assumes that attackers could passively listen and accurately determine the current \gls{csi} of normal users, without investigating how to attack the \gls{drl}-based multi-user scheduling system.

        In conclusion, our work differs from existing literature by assume that adversaries only manipulate a small fraction of the attacker's input to \gls{drl} models, without  the knowledge of local observations of normal users.

    \section*{Conclusion\tvn{\st{Why not Conclusion?}}}

In this paper, we present \gls{fggm}, a technique designed to generate adversarial observations that compromise the functionality of \gls{drl}-based resource allocation algorithms. 
Unlike other previous attacking schemes, \gls{fggm} only controls a part of the observations, and does not require the exact knowledge of the other uncontrollable parts. 
Experimental results on \gls{drl}-based scheduler for massive \gls{mumimo} networks show that adversarial observations generated by \gls{fggm} effectively reduced the allocated resources to victims without the exact knowledge of their local observations. 
By constructing an objective function based on output bounds, the adversarial inputs have been optimized using stochastic gradient descent to achieve specific goals, such as minimizing the likelihood of the victim being selected for transmission. 

\tvn{can?  it did and you have experiments to show that}
\tvn{Where's the discussion?}
\tvn{just a case study? the reviewer might say then may be you should do more experiments?}
\tvn{as a reviewer I would say why don't you do it and get those results and show us?}
\tvn{Give some suggestion on what kind of loss functions and other exploitation techniques here}

\thanh{\section*{Comments}}

\kei{\subsection{Kei sensei}}
    \kei{Introduction, related works, general threat model and threat model, specific system model, threat model, and attack. Experiment.}
    
    \kei{cite the NCC paper is not very convincing}
    
    \kei{RL can solve multiple problem. Contribution on learning part. Start from the RL can solve multiple problem first then the particular scenario later. Like the TMC, focus on attack the learning. }

\tvn{\subsection{Vu}}
    \tvn{BS is a weird acronym}
    \thanh{It is normal in communication society, I deleted them anyway, as there are too many acronyms.}
    
    \tvn{too much text on related work and little txt on the technique proposed by this work.  I would shorten related work and add a paragraph on this work, its novelty, etc.}
    \thanh{I removed the text on related work and focus on the assumptions of my target system.}

    \tvn{Also, I don't clearly see the main idea in this intro.  Should write about it and emphasis on what the main idea is (which implicitly shows how it can address issues of existing work). For example if it is about the formulation to a DNN verification which allows the application SOTA DNN tools to solve the problem then write about it. I would also mention/summarize significant evaluation results.}
    \thanh{Done, but not sure you would be satisfied. Please help me to confirm the revision is enough. }
    
    \tvn{You also did evaluation and case studies to show various RQs.  These are also contributions.}
    \thanh{I have changed that}
    
    \tvn{Do you plan to release tools, prototypes, benchmarks, etc?  These are contributions.}
    \thanh{I have added that to the contribution. }

    \tvn{Too many acronyms. BS to base station. }    
    \thanh{I removed many acronyms. }    

    \tvn{RQ1:too short. Summarize (by giving some numbers) the findings above here.}

    \tvn{Don't talk too much about Marabou, DPLL, SMT, and Polytopes and such.
         Not a focus of this work, which use them as blackbox.
         Instead, briefly talk about major accomplishment, e.g., sota tool such as neuralsat can efficiently prove/disprove properties of networks with million of parameters.}
    \hai{I merged this part with preliminaries, and proposed method}

    \tvn{could be different community has diff preference. But this Related work section is too long for me. I would keep it less than a page.}
    \thanh{TODO}
    
    \tvn{In the beginning of evaluation "investigate the conditions ..." this doesn't sound interesting.  It also doesn't sound "enough",  e.g., why just eval these conditions? what about the evaluation of the technique itself?}
    \thanh{I have attempted to fix it.}

\hai{\subsection{Hai}}
    \hai{In the introduction, focus on introduce problem 50\%, discuss the proposal 50\%}
    \thanh{done}
    
    \hai{Cut these related works. }
    \hai{This part (the related work) should be more compacted}
    \thanh{TODO}
    
    \hai{Chien: Conceptual contribution is what?}
    \thanh{done}
    
    \hai{The contribution should be merged above. The enumerated list of contribution should be short to highlight. }
    \thanh{done}

\thanh{\subsection{Thanh's todo list}}
    \thanh{\st{Convert all greybox to grey-box.}}
    
    \thanh{\st{Convert all overapproximat to over-approximat.}}
    
    \thanh{\st{Ensure the percentage are correct up to 1 digit.}}

    \thanh{Redraw figures for OptMR to OptCI, and boxplot to barchart as JFI calculated at the end only. }
    
\john{\subsection{John}}

    \printbibliography
\end{document}